\documentstyle[epsfig]{mn}

\begin{document}

\title[Relativistic Accretion Disc Spectra]%
{General relativistic spectra of accretion discs around rapidly
rotating neutron stars: Effect of light bending}

\author[Bhattacharyya et al.]
{Sudip Bhattacharyya$^{1,2,\ast}$, 
Dipankar Bhattacharya$^{3,\ast}$ and 
Arun V. Thampan$^{4,\ast}$\\ 
$^1$Joint Astronomy Program, Indian Institute of Science, 
Bangalore 560012, India\\
$^2$Indian Institute of Astrophysics, 
Bangalore 560034, India\\
$^3$Raman Research Institute, Bangalore 560012, 
India\\
$^4$Inter-University Centre for Astronomy and 
Astrophysics, Pune 411007, India\\
$^{\ast}$e-mail: sudip@physics.iisc.ernet.in,dipankar@rri.res.in,arun@iucaa.ernet.in}

\date{26 Dec 2000}

\maketitle

\begin{abstract}
We present computed spectra, as seen by a distant observer, 
from the accretion disc around a rapidly rotating neutron star. 
Our calculations are carried out in a fully general relativistic 
framework, with exact treatment of rotation.  We take into 
account the Doppler shift, gravitational redshift and light 
bending effects in order to compute the observed spectrum.  
We find that light bending significantly modifies the high-energy 
part of the spectrum.  Computed spectra for slowly rotating neutron 
stars are also presented.  These results would be important for 
modelling the observed X-ray spectra of Low Mass X-ray Binaries 
containing fast spinning neutron stars.
\end{abstract}


\section{Introduction}

The central accretors in a large number of low mass X-ray binaries 
(LMXB) are believed to be neutron stars, rotating rapidly due to 
accretion-induced angular momentum transfer. LMXBs are thought 
to be the progenitors of milli-second (ms) radio pulsars 
(Bhattacharya \& van den Heuvel 1991) like PSR~1937+21 with 
$P \sim 1.56$ ms (Backer et al 1982).  The recent discovery 
of ms ($P \sim 2.49$ ms) 
X-ray pulsations in XTE~J1808-369 (Wijnands \& van der Klis 1998) 
has strengthened this hypothesis.  Kilohertz Quasi-Periodic 
Oscillations (kHz QPOs) seen in a number of LMXBs is another 
indicator of the rapid rotation of the accreting neutron star.  
For example, in beat-frequency models of kHz QPO, the difference 
($\sim 300-500$~Hz) in the frequencies of two simultaneously 
observed kHz QPO peaks is interpreted as the rotational 
frequency of the neutron star (e.g. van der Klis 2000). 

Fast rotation makes the neutron star considerably oblate and 
also modifies, both the interior and exterior  space-time 
geometry. These, 
in turn, modify the size and the temperature profile of the 
accretion disc and hence the corresponding spectrum 
(Bhattacharyya et al. 2000;2001). General relativity also 
plays an important role in shaping the spectrum. As neutron 
stars are very compact objects, the effect of general relativity 
is very significant, particularly near the surface of the star. 

For luminous LMXBs, the observed X-ray spectrum can be well 
fitted by the sum of a multi-colour blackbody spectrum (presumably
from the accretion disc) and a single temperature blackbody 
spectrum (presumably from the boundary layer) 
(see Mitsuda et al. 1984). The multi-colour spectrum can be 
calculated if the temperature profile in the accretion disc 
is known. Recently Bhattacharyya et al. (2000) have calculated 
the disc temperature profile as a function of the spin rate 
($\Omega$) and the mass ($M$) of the neutron star for different 
equations of state (EOS), including the effects of rotation and 
general relativity. In this paper, we calculate the observed 
spectrum from the accretion disc, using the temperature profiles. 
While doing so, we take into account the effect of gravitational 
redshift, Doppler broadening and the light-bending effect in the 
gravitational field of the accreting star.

For non-rotating neutron stars, an attempt has been made by 
Ebisawa et al. (1991) to compute the disc spectrum including 
the effects mentioned above.  Sun \& Malkan (1989) included
relativistic effects of disc inclination, including Doppler
boosting, gravitational focussing, and gravitational redshift,
on the observed disc spectra for both Kerr and Schwarzschild
black holes (in the context of supermassive black holes).
They find (as we also do) that the higher energy
part of the spectrum gets significantly modified due to the
general relativisitic effects.  A computation similar to that
of Ebisawa et al. (1991), for Galactic black hole candidates,
has been done by Asaoka (1989) using the
Kerr metric.  However, our work incorporating both: 
the full general relativistic effects of rapid rotation, 
as well as realistic equations of state describing neutron star
interiors, using an appropriate metric is the first
calculation for rotating neutron stars.  In the present
work we ignore the effects of the stellar magnetic field, so
our results are applicable to weakly magnetised neutron stars.

The structure of the paper is as follows. In section 
2, we describe a framework for the calculation of neutron 
star structure, disc temperature profile and observed spectrum. 
We also comment here on the the chosen equations of state. We 
describe the numerical procedure for spectrum calculation in 
section 3. The results of the spectrum calculation are displayed 
in section 4. We  summarise the conclusions in section 5 and 
enlist some relevant mathematical expressions in the Appendix. 
 
\section{Theory}

\subsection{Neutron star structure calculation}

We calculate the structure of a rapidly rotating neutron star
for realistic EOS models, using the same procedure as Cook et al. 
(1994) (see also Datta et al 1998). With the assumption that 
the star is rigidly rotating and a perfect 
fluid, we take a stationary, axisymmetric, asymptotically flat 
and reflection-symmetric (about the equatorial plane) metric, 
given by
\begin{eqnarray}
dS^2 & = & g_{\rm \mu\nu} dx^{\rm \mu} dx^{\rm \nu} 
(\mu, \nu = 0, 1, 2, 3) 
\nonumber\\
 & = & -e^{\rm {\gamma + \rho}} dt^2 + e^{\rm {2\alpha}} 
(d{\bar r}^2 + {\bar r}^2 d {\theta}^2) \nonumber \\
 &  & 
 + e^{\rm {\gamma - \rho}} {\bar r}^2 \sin^2\theta 
 {(d\phi - \omega dt)}^2 \label{eq:metric}
\end{eqnarray}
where the metric potentials $\gamma, \rho, \alpha$ and the 
angular speed ($\omega$) of zero-angular-momentum-observer 
(ZAMO) with respect to infinity, are all functions of the 
quasi-isotropic radial coordinate ($\bar r$) and polar angle 
($\theta$). The quantity $\bar r$ is related to the 
Schwarzschild-like radial coordinate ($r$) by the equation 
$r = \bar r e^{\rm {(\gamma - \rho)/2}}$. 
We use the geometric units $c = G = 1$ in this paper.

We solve Einstein's field equations and the equation of 
hydrostatic equilibrium self-consistently and numerically 
from the centre of the star upto infinity 
to obtain $\gamma, \rho, \alpha, \omega$ and $\Omega$ 
(angular speed of neutron star with respect to infinity) 
as functions of $\bar r$ and $\theta$. This is done for a 
particular EOS and assumed values of central density and
ratio of polar to equatorial radii. The obtained numerical 
equilibrium solutions of the metric enable us to calculate 
bulk structure parameters, such as gravitational mass ($M$), 
equatorial radius ($R$), angular momentum ($J$) etc. of the 
neutron star. We also calculate the radius ($r_{\rm orb}$) 
of the innermost stable circular orbit (ISCO), specific 
energy ($\tilde E$) and specific angular momentum ($\tilde l$) 
of a test particle in a Keplerian orbit and the Keplerian 
angular speed ($\Omega_{\rm K}$) (see Thampan \& Datta 1998 
for a detailed description of the method of calculation).

\subsection{Equation of state}

The neutron star structure parameters are quite sensitive to 
the chosen EOS.  In the literature, there exist a large number of 
EOS ranging from very soft to very stiff. For the purpose 
of a general study, we have chosen four EOS, of which one is 
soft (EOS : A, Pandharipande (1979) (hyperons)), one is 
intermediate (EOS : B, Baldo, Bombaci \& Burgio (1997) (AV14+3bf))
in stiffness and two are stiff (EOS : C, Walecka (1974) and EOS : D, Sahu, 
Basu \& Datta (1993)). Of these EOS : D is the stiffest. 

The structure of a neutron star for a given EOS is described 
uniquely by two parameters : the  gravitational mass ($M$) and 
the angular speed ($\Omega$).  For each chosen EOS, we construct 
constant $M$ equilibrium sequences with $\Omega$ varying from the 
non-rotating case upto the centrifugal mass shed limit (rotation
rate at which inwardly directed gravitational forces are balanced
by outwardly directed centrifugal forces). 
Depending on $\Omega$, $M$  and the EOS model,
neutron stars may have $R>r_{\rm orb}$ or $R<r_{\rm orb}$.

\subsection{Disc temperature profile calculation}

The effective temperature of a thin blackbody disc is given by
\begin{eqnarray}
T_{\rm eff}(r) & = & (F(r)/\sigma)^{1/4} \label{eq:Teff}
\end{eqnarray}
where $\sigma$ is the Stefan-Boltzmann constant and $F$ is 
the X-ray energy flux per unit surface area. We calculate 
$F$ using the expression of  Page \& Thorne (1974) (valid 
for geometrically thin non-self-gravitating disc embedded 
in a general axisymmetric space--time) for the accretion 
disc around a rotating blackhole : 
\begin{eqnarray}
F(r) & = & \frac{\dot{M}}{4 \pi r} f(r)\label{eq:Fr}
\end{eqnarray}
where 
\begin{eqnarray}
f(r) & = & -\Omega_{{\rm K},r} (\tilde{E} - \Omega_{\rm K} 
\tilde{l})^{-2} \int_{r_{\rm in}}^{r} (\tilde{E} - \Omega_{\rm K} 
\tilde{l}) \tilde{l}_{,r} dr \label{eq:fr}
\end{eqnarray}
Here $r_{\rm in}$ is the disc inner edge radius and a 
comma followed by a variable as subscript to a quantity, 
represents a derivative of the quantity with respect to 
the variable. 

Bhattacharyya et al. (2000) argue that, though formulated 
for the case of black holes, these expressions hold quite 
well also for the case of neutron stars.  If $R > r_{\rm orb}$, 
then the disc will touch the star and we take 
$r_{\rm in} = R$.  Otherwise, $r_{\rm in} = r_{\rm orb}$. 

The temperature profile, of the accretion disc calculated as 
described in this and the previous subsections is a function 
of $M$ and $\Omega$ of the central star for any adopted EOS. 

\subsection{Calculation of the spectrum}

For calculating the disc spectrum in full general relativity including 
light bending effect for rapid rotation we adopt the following procedure.

The observed spectrum from the accretion disc is given by
\begin{eqnarray}
F(E_{\rm ob}) & = & (1/E_{\rm ob})\int I_{\rm ob}(E_{\rm ob}) 
d\Pi_{\rm ob} \label{eq:FEob}
\end{eqnarray}
where the subscript `ob' denotes the quantity in observer's 
frame, $F$ is expressed in photons/sec/cm$^2$/keV, $E$ is 
photon energy in keV, $I$ is specific intensity and $\Pi$ 
is the solid angle subtended by the source at the observer.

As $I/E^3$ remains unchanged along the path of a photon 
(Misner et al. 1973), it is possible to calculate 
$I_{\rm ob}$, if $I_{\rm em}$ is known (we use the 
subscript `em' to denote the quantities in the emitter's frame). 
We assume the accretion disc to radiate like a diluted 
blackbody (see Shimura \& Takahara, 1995). So $I_{\rm em}$ is
given by
\begin{eqnarray}
I_{\rm em} & = & (1/f^4) B(E_{\rm em},T_{\rm c}) \label{eq:Iem}
\end{eqnarray}
where $f$ is the colour factor of the disc, $B$ is Planck 
function and $T_{\rm c}$ = $f T_{\rm eff}$.

The quantities $E_{\rm ob}$ and $E_{\rm em}$ are related 
by $E_{\rm em}$ = $E_{\rm ob} (1 + z)$, where (1 + $z$) 
contains the effects of both gravitational redshift
and Doppler shift. (1 + $z$) is given by (Luminet 1979)
\begin{eqnarray}
1 + z & = & (1 + \Omega_{\rm K} b \sin \alpha \sin i) 
(-g_{tt} - 2 \Omega_{\rm K} g_{t \phi} \nonumber \\  
 &   & - \Omega_{\rm K}^2 g_{\phi\phi})^{-1/2}  \label{eq:1pz}
\end{eqnarray}
where $i$ is the inclination angle of the disc with respect to
an observer at infinity, 
$b$ is the impact parameter of the photon relative to the 
line joining the source and the observer and $\alpha$ is 
the polar angle of the position of the photon on the 
observer's photographic plate. Therefore $b\,db\,d\alpha$ 
is the apparent area of a disc element at observer's sky 
and the corresponding solid angle is given by 
\begin{eqnarray}
d\Pi_{\rm ob} & = & b\;db\;d\alpha \over D^2 \label{eq:Piob}
\end{eqnarray}
where $D$ is the distance of the source from the observer.

\section{Numerical procedure}

 For a configuration, described by $M$ and $\Omega$ (and thus
 specified by a set of $g_{\mu \nu}$), we obtain $\Omega_{\rm K}$.
 To calculate the spectrum for a given value of $i$ with light
 bending effects, we backtrack the photons path from the observer
 to the disc, using standard ray tracing techniques
 (e.g. Chandrasekhar 1983) and the relevant boundary conditions.
 For the metric (\ref{eq:metric}) that we use, the equations
 of motion for photons are provided in the Appendix.  We cover
 the disc between radii $r_{\rm in}$ and
 $r_{\rm mid}=1000 r_{\rm g}$;  $r_{\rm in}$ being the radius
 of the inner edge of the disc and $r_{\rm g}$ the Schwarzschild
 radius (increasing $r_{\rm mid}$ has no significant effect
 on the spectrum).
 Beyond $r_{\rm mid}$, we ignore the
 effect of light-bending i.e., we take
 $b \sin\alpha = r \sin\phi$ ($\phi$ is the azimuthal
 angle on disc plane) and
 $d\Pi_{\rm ob}$ = $(r\;dr\;d\phi\;\cos i)\;/\;D^2$
 (Bhattacharyya et al. 2001).

We have performed several consistency checks on our results:
(1) by switching off the light bending effect (i.e. by considering 
    flat space-time while backtracking the photon's path), we see 
    that the spectrum matches very well with that computed by ignoring 
    light bending effects (calculated by an independent code --
    Bhattacharyya et al. 2001).   Also, in this case,
    the analytically calculated values of several quantities 
    on the disc plane (e.g. $r$, $\phi$, $d\phi/dt$, $d\theta/dt$ 
    etc.) are reproduced satisfactorily by our numerical method, 
(2) an increase in the number of grid points on the ($b$,$\alpha$) 
    plane do not have any significant effect on the computed spectrum,
(3) the spectrum matches very well with the Newtonian spectrum 
    (Mitsuda et al. 1984) at low energy limit.  
This would imply that for higher frequencies, our spectrum is 
correct to within 0.2\% to 0.3\%. 

\section{Results}

We calculate the general relativistic spectrum from the accretion 
disc around rapidly rotating neutron star, taking into account the 
light bending effect.  The spectrum is calculated as a function 
of 6 parameters : $M$, $\Omega$, distance of the source ($D$), 
inclination angle ($i$) (for face-on, $i = 0\degr$), 
accretion rate ($\dot M$) and colour factor $f$, 
for each of the chosen EOS. Our results are displayed 
in Figs. 1 to 5.  In all the displayed spectra, we have 
assumed $M = 1.4 M_{\odot}$ (canonical mass for neutron stars), 
$D = 5$ kpc and $f = 2.0$.  

In Fig. 1, we have plotted the Newtonian spectrum and 
GR spectra with (LBGR) and without (NLBGR) light bending 
effect, keeping the values of all the parameters same. At 
10 keV, the Newtonian flux is almost 2.5 times the LBGR flux. 
This is quite expected, because in the inner parts of the disc, 
Newtonian temperature is considerably higher than the GR temperature 
(see Fig. 2 of Bhattacharyya et al., 2000). LBGR flux is about 
50\% higher than NLBGR flux at 10 keV. This is because light bending 
causes the disc to subtend a larger solid angle at the observer than 
otherwise. Thus the general effect of light bending is to increase 
the observed flux.  

\begin{figure}
\epsfig{file=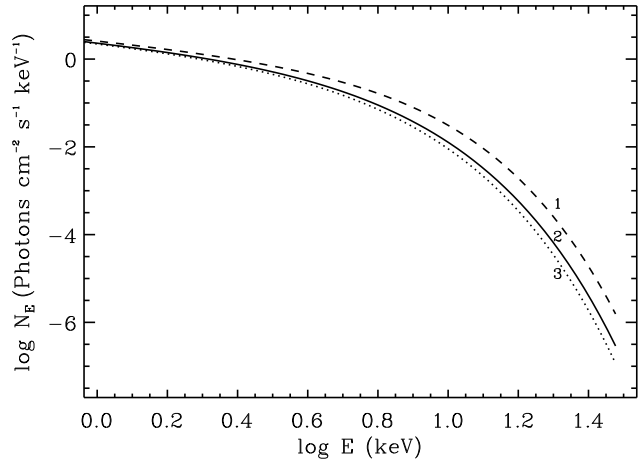}
\caption{Effect of general relativity: spectra from accretion 
disc around a neutron star of mass $1.4 M_\odot$. All the 
curves are for EOS model (B), $\Omega = 0$, $D = 5$~kpc,
$i = 60\degr$, $\dot M = 10^{18}$~g/sec and $f = 2$. 
Curve (1) corresponds to the Newtonian case, 
curve (2) to the general relativistic case including the
effect of light bending and 
curve (3) to the general relativistic case without considering 
the effect of light bending.}
\end{figure}

According to Shimura \& Takahara (1995), the thin blackbody 
description of the accretion disc, as adopted in this paper, 
is valid for $0.1 {\dot M}_{\rm E} < \dot M < {\dot M}_{\rm Edd}$, 
where ${\dot M}_{\rm E} \equiv L_{\rm Edd}/c^2$. Here $L_{\rm Edd}$ 
is the Eddington luminosity and ${\dot M}_{\rm Edd}$ is the true 
Eddington accretion rate.  For the purpose of demonstration,
 we have taken three different values of $\dot M$ in this range 
(for the mass shed configuration) and plotted the corresponding 
spectra in Fig. 2. As is expected, we see that the high energy 
part of the spectrum is more sensitive to the value of $\dot M$. 
It is seen that the spectra for different values of $\dot M$ are 
easily distinguishable. 

\begin{figure}
\epsfig{file=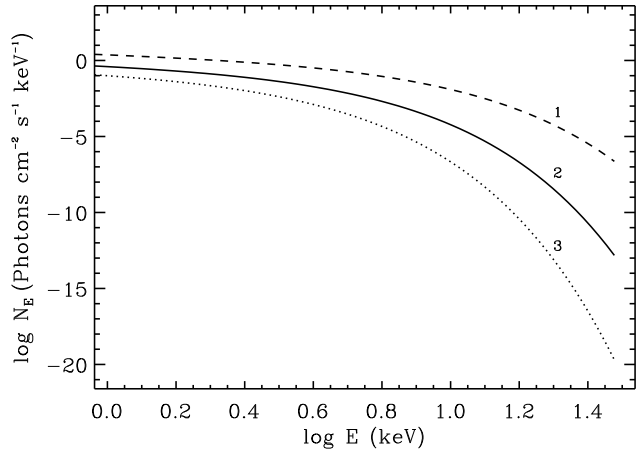}
\caption{Accretion rate dependence: general relativistic 
spectra including light bending effects from accretion 
disc around a neutron star of mass shed limit configuration 
($\Omega = 7001$~rad/s).  
Curve (1) corresponds to $\dot M = 10^{18}$~g/sec, 
curve (2) to $\dot M = 10^{17}$~g/sec and 
curve (3) to $\dot M = 2 \times 10^{16}$~g/sec. 
The values of all the other parameters are as in Fig. 1.}
\end{figure}

The inclination angle $i$ is a very important parameter in 
determining the shape of the spectrum and its overall normalisation. 
In Fig. 3, we have plotted the spectra for four inclination angles, 
for the mass shed configuration. We see that the observed flux at low 
energies is higher for lower values of $i$.  This is due to 
the projection effect (proportional to $\cos i$). But at higher 
energies ($> 10$~keV) this trend is reversed mainly because Doppler 
effect becomes important. The most energetic photons largely come 
from the inner portion of the disc, where the linear speed of 
accreted matter is comparable to the speed of light. The net 
effect of Doppler broadening is a net blue shift of the 
spectrum, as a larger amount of flux comes from the 
blue-shifted regions than from the red-shifted regions. 
This is a monotonic trend, but it will be noticed from Fig.~3 
that the curve for $i=85\degr$ overcomes that for 
$i=60\degr$ only at the edge of the figure, {\em i.e.} at
energies $\geq 30$~keV.  This is due to the fact that
between these two inclinations the difference in the $\cos i$
factor is severe, and the blueshift overcomes this only at
high energies. 

\begin{figure}
\epsfig{file=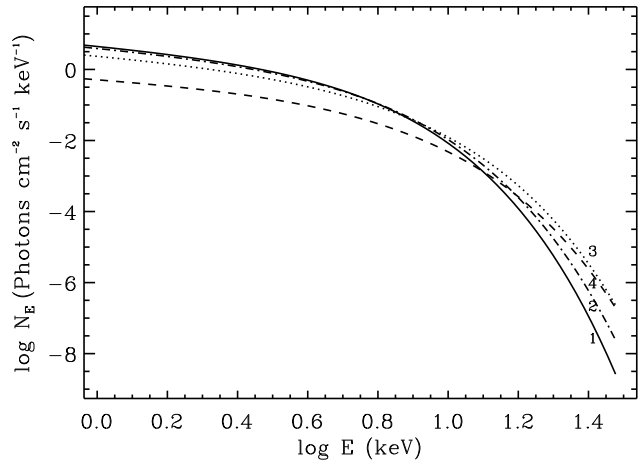}
\caption{Inclination angle dependence: general relativistic 
spectra including light bending effects from accretion disc 
around a neutron star of mass shed limit configuration 
($\Omega = 7001$~rad/s). 
Curve (1) corresponds to $i = 0\degr$, 
curve (2) to $i = 30\degr$, 
curve (3) to $i = 60\degr$ and 
curve (4) to $i = 85\degr$. 
The values of all the other parameters are as in Fig. 1.}
\end{figure}
 
In Fig. 4, we have four panels for four inclination 
angles. In each panel, we have shown spectra for 3 different 
$\Omega$ (corresponding to non-rotating, intermediate and the 
mass shed configurations).  With the increase of $\Omega$, disc 
temperature profile does not vary monotonically (see Fig. 3a 
of Bhattacharyya et al., 2000). Hence the behaviour of the 
spectrum is also non-monotonic with $\Omega$. For non-rotating 
and mass shed configurations (for the assumed values of other 
parameters) the temperature profiles are very similar.  
As a result, the plotted spectra for these two cases lie
almost on top of each other.  However, for $i = 0\degr$
the flux corresponding to the mass shed configuration
is slightly higher than that for $\Omega=0$, while
the case is opposite at higher inclinations.  This is a
result of the inclination dependence of the $(1+z)$ factor
given in Equation~(\ref{eq:1pz}).

\begin{figure}
\epsfig{file=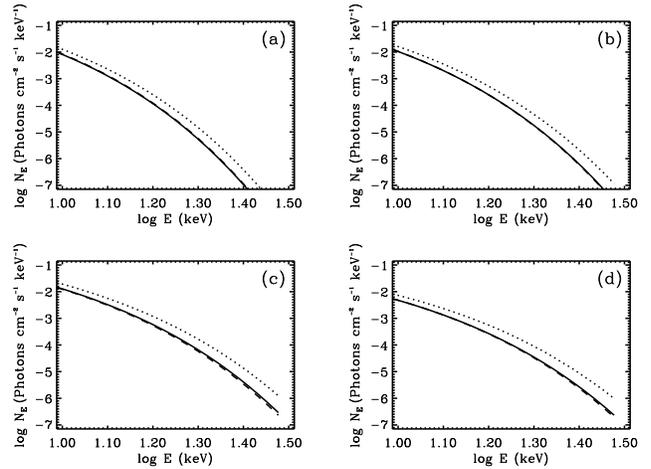}
\caption{Rotation rate dependence: general relativistic spectra
including light bending effects from accretion disc around a 
neutron star. 
Panel (a) corresponds to $i = 0\degr$, 
panel (b) to $i = 30\degr$, 
panel (c) to $i = 60\degr$ and 
panel (d) to $i = 85\degr$. 
In each panel, the solid curve corresponds to $\Omega = 0$ rad/s, 
the adjacent dashed curve corresponds to $\Omega = 7001$ rad/s 
(the mass shed limit) 
and the dotted curve corresponds to $\Omega = 3647$ rad/s.
The values of all the other parameters are as
in Fig. 1.}
\end{figure}

In Fig. 5, we have compared the spectra for the four 
EOS models adopted by us, for configurations at the 
respective mass-shed limits (which correspond to different 
values of $\Omega$ because of the EOS dependence of the stellar
structure).  The values of all other parameters have been kept 
the same.  We see that the total flux received varies 
monotonically with the stiffness parameter, and is higher for 
the softer EOS.  This effect has been noticed earlier by 
Bhattacharyya et al. (2000). We see that at high energies
the fluxes for different EOS are considerably different. 
Therefore, fitting the observed spectra of LMXBs with our 
model spectra, particularly in hard X-rays, may provide a 
way to constrain neutron star EOS.  Of course it must be
remembered that these computations have been made assuming 
that the magnetic field of the compact object does not 
limit the inner boundary of the accretion disc.  In the
presence of a sufficiently strong magnetic fields, 
appropriate modifications must be taken into account 
while calculating the expected flux at high energies.  

\begin{figure}
\epsfig{file=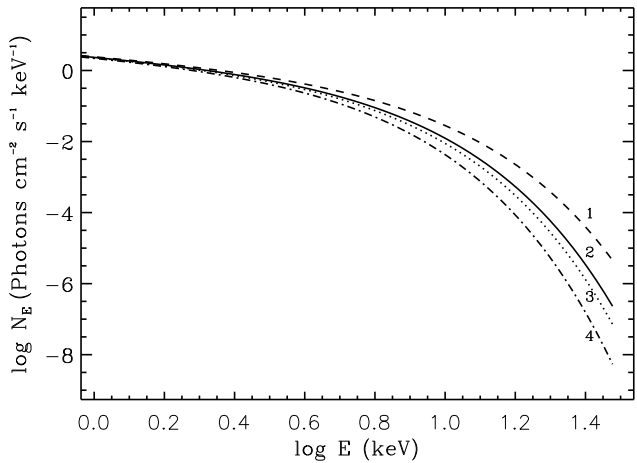}
\caption{EOS model dependence: general relativistic spectra 
including light bending effects from accretion disc
around a neutron star of mass shed configuration. 
Curve (1) corresponds to the EOS model (A) ($\Omega = 11026$~rad/s), 
curve (2) to the EOS model (B) ($\Omega = 7001$~rad/s), 
curve (3) to the EOS model (C) ($\Omega = 6085$~rad/s) and 
curve (4) to the EOS model (D) ($\Omega = 4652$ rad/s). The values 
of all the other parameters are as in Fig. 1.}
\end{figure}

\section{Conclusion}
In this paper we have computed the observed radiation 
spectrum from accretion discs around rapidly rotating 
neutron stars using fully general relativistic disc 
models.  This is the first time such a calculation has 
been made in an exact way, without making any approximation 
in the treatment of either rotation or general relativity.  
In computing the observed spectrum from the disc, we 
explicitly include the effects of Doppler shift, 
gravitational redshift and light-bending for an appropriate
metric describing space--time around rapidly rotating 
neutron stars.  We find that the effect of 
light bending is most important in the high-energy ($> 3$~keV) 
part of the observed spectrum.  Photons at these high energies 
originate close to the central star, and hence their trajectories 
are most affected by the light-bending effect.  Depending on the 
viewing angle, this can enhance the observed flux at $\sim 10$~keV 
by as much as $250$\% compared to that expected if light-bending 
effects are neglected.  

The calculations presented here deal only with the multicolour 
blackbody disc.  In reality, of course, there will be additional
contributions to the observed spectrum from the boundary layer
and also possibly from an accretion disc corona, both of which 
are likely to add a power-law component at high energies 
(Popham \& Sunyaev 2000, Dove {\it et al} 1997).  On the other 
hand, the spectra presented in Figs (2), (3) and (5) should 
remain essentially unaffected by boundary layer contribution, 
as these are for neutron stars rotating near the mass shed 
limit for which the boundary layer luminosity will be negligible.  
For slowly rotating neutron stars, the disc component of the 
spectrum can be obtained by fitting and removing the contribution 
of the boundary layer, provided a good model for the boundary 
layer spectrum is available.  Popham \& Sunyaev (2000) have 
calculated the boundary layer spectrum in the Newtonian 
approximation.  General relativistic
modifications need to be included in these calculations to get
a realistic estimate of the spectrum of the boundary layer.  We
plan to address this issue in a future work.
In the slow rotation case, the spectrum of 
the disc itself may be somewhat modified by the presence of a
boundary layer if it extends beyond the disc inner radius assumed in
our computations here, thus curtailing the inner edge of the disc.

In addition to the contribution of the boundary layer, the
possible contribution of an accretion disc corona to the emergent
spectrum could also be significant. To be able to constrain
the EOS models of neutron stars using the observed spectrum,
this contribution must also be accurately estimated.  
In the present work, we restrict ourselves to thin
black--body and non--magnetic accretion discs in order to
understand the effect of the EOS models describing neutron stars
on the spectrum of the accretion disc alone (and thus neglect the 
effect of a corona). We view this as the first step in modelling 
of spectra of accreting 
neutron stars including the effects of general relativity and
rotation.  

The comparison of the non-rotating limit of our results with 
those of the fitting routine {\it GRAD} in the X--ray spectral 
reduction package {\it XSPEC} (Ebisawa et al. 1991), shows 
that the latter model overpredicts the high-energy 
component of the flux by a large factor.  
With the help of Ebisawa \& Hanawa 
({\it private communication}) we have been able
to trace this disagreement to certain simplifying approximations 
made in the {\it GRAD} code, as well as a couple of incorrect 
expressions being used there.
Conclusions based on the use of the {\it GRAD} routine may 
therefore need to be revised in the light of the new calculations 
presented here.

The computation of the complete spectrum in the manner 
presented here is rather time-consuming and therefore not 
quite suited to routine use.  Therefore, in order to make 
our results available for routine spectral 
fitting work, we intend to present a series of approximate 
parametric fits to our computed spectra in a forthcoming 
publication.

The spectra presented here will find use in
constraining the combined parameter set of the mass, the 
rotation speed and, possibly, the EOS, particularly of 
weakly magnetised, rapidly rotating neutron stars.  The 
relevant signatures are most prominent in hard X-rays, 
above $\sim 10$~keV.  Sensitive observations of hard-X-ray 
spectra of LMXBs, therefore, are needed to fully utilise the 
potential of these results.  

\section*{acknowledgements}
We would like to thank late Bhaskar Datta for the discussions at 
the early stage of the work. We thank Ranjeev Misra for providing 
us with the {\it GRAD} routine and for many 
discussions. 
We are very thankful to K. Ebisawa and T. Hanawa for discussions 
which helped resolve the discrepancy between our results 
and those obtained from the {\it GRAD} routine.
We are also grateful to B.R. Iyer for valuable suggestions 
during the course of this work.
SB thanks Pijush Bhattacharjee for encouragement.

\section*{Appendix}

For convenience, we use $\mu$ ($= \cos\theta$) instead of $\theta$ and $s$
(= $\bar r/(A + \bar r$)) instead of $r$ as the coordinates. Consequently, 
the equation (\ref{eq:metric}), becomes 
\begin{eqnarray}
dS^2 & = & -e^{\rm {\gamma + \rho}} dt^2 + e^{\rm {2\alpha}} ((A^2/(1-s)^4) 
ds^2 \nonumber \\
 &   & + A^2 (s/(1-s))^2 (1/(1-\mu^2)) d\mu^2) \nonumber \\
 &   & + e^{\rm {\gamma - \rho}} 
A^2 (s/(1-s))^2 (1-\mu^2) (d\phi - \omega dt)^2
\end{eqnarray}
Here, $A$ is a known constant of the dimension of distance.

Now it is quite straight forward to calculate the geodesic equations for 
photons, which are given below.
\begin{eqnarray}
dt/d\lambda & = & e^{\rm {-(\gamma + \rho)}} (1 - \omega L)
\end{eqnarray}
\begin{eqnarray}
d\phi/d\lambda & = &  e^{\rm {-(\gamma + \rho)}} \omega (1 - \omega L) \nonumber \\ 
 &   & + L/(e^{\rm {\gamma - \rho}} A^2 (s/(1-s))^2 (1-\mu^2))
\end{eqnarray}
\begin{eqnarray}
(ds/d\lambda)^2 & = &  e^{\rm {-2\alpha}} ((1-s)^4/A^2) 
(e^{\rm {-(\gamma + \rho)}} (1 - \omega L)^2 \nonumber \\ 
 &   & - L^2/(e^{\rm {\gamma - \rho}} 
A^2 (s/(1-s))^2 (1-\mu^2))) \nonumber \\ 
 &   & - s^2 (1-s)^2 (1/(1-\mu^2)) y^2
\end{eqnarray}
\begin{eqnarray}
d\mu/d\lambda & = &  y
\end{eqnarray}
\begin{eqnarray}
dy/d\lambda & = & -2 (\alpha_{\rm ,s} + (1/(s(1-s)))) y (ds/d\lambda) \nonumber \\ 
 &   & + \alpha_{\rm ,\mu} (1/(s(1-s)))^2 (1-\mu^2) (ds/d\lambda)^2 \nonumber \\ 
 &   & - (\alpha_{\rm ,\mu} + (\mu/(1-\mu^2)) y^2 \nonumber \\
 &   & + ((1/2) e^{\rm {\gamma - \rho - 2\alpha}} (\gamma_{\rm ,\mu} - 
\rho_{\rm ,\mu}) (1-\mu^2)^2 \omega^2 \nonumber \\ 
 &   & - e^{\rm {\gamma - \rho - 2\alpha}} \mu (1-\mu^2) \omega^2 \nonumber \\  
 &   & + e^{\rm {\gamma - \rho - 2\alpha}} (1-\mu^2)^2 \omega \omega_{\rm ,\mu}
 \nonumber \\ 
 &   & - (1/2) e^{\rm {\gamma + \rho - 2\alpha}} (\gamma_{\rm ,\mu} + 
\rho_{\rm ,\mu}) \nonumber \\  
 &   & ((1-s)/(A s))^2 (1-\mu^2)) (dt/d\lambda)^2 \nonumber \\
 &   & + e^{\rm {\gamma - \rho - 2\alpha}} (1-\mu^2)^2 (-\omega_{\rm ,\mu} 
- \omega (\gamma_{\rm ,\mu} - \rho_{\rm ,\mu}) \nonumber \\ 
 &   & + 2 \omega (\mu/(1-\mu^2))) (d\phi/d\lambda) (dt/d\lambda) \nonumber \\ 
 &   & + e^{\rm {\gamma - \rho - 2\alpha}} (1-\mu^2)^2 ((1/2) 
(\gamma_{\rm ,\mu} - \rho_{\rm ,\mu}) \nonumber \\ 
 &   & - \mu/(1-\mu^2)) (d\phi/d\lambda)^2
\end{eqnarray}
where, $\lambda$ is the affine parameter, $L$ is the negative of the ratio of 
the $\phi$-component of the angular momentum and the $t$-component of the 
angular momentum of photon and a comma followed by a variable as subscript to 
a quantity, represents a derivative of the quantity with respect to the variable.


\begin{thebibliography}{}

\bibitem[]{} Asaoka, I. 1989, {\it Pub. Astr. Soc. Japan}, {\bf 41},
763.

\bibitem[]{} Backer, D.C., Kulkarni, S.R., Heiles, C., 
             Davis, M.M., Goss, W.M. 1982, Nature, 300, 615

\bibitem[]{} Baldo, M., Bombaci, I., Burgio, G.F. 1997, {\it Astr.
Astrophys.}, {\bf 328}, 274.

\bibitem[]{} Bhattacharya, D., \& van den Heuvel, E.P.J. 1991, 
Phys. Repts., {\bf 203}, 1.

\bibitem[]{} Bhattacharyya, S., Misra, R., \& Thampan, A.V. 2001, 
ApJ, in press.

\bibitem[]{} Bhattacharyya, S., Thampan, A.V., Misra, R., \& Datta, B.
2000, {\it Astrophys. J.}, 542, 473

\bibitem[]{} Chandrasekhar, S. 1983, {The Mathematical Theory of Black
           Holes}, Oxford University Press, London

\bibitem[]{} Cook, G.B., Shapiro, S.L., \& Teukolsky, S.A. 1994, 
{\it Astrophys. J.}, {\bf 424}, 823.

\bibitem[]{} Dove, J.B., Wilms, J., Begelman, M.C. 1997, {\it 
Astrophys. J.}, {\bf 487}, 747.

\bibitem[]{} Liang, E.P.T., \&  Price, R.H. 1977,  {\it 
Astrophys. J.}, {\bf 218}, 247.

\bibitem[]{} Ebisawa, K., Mitsuda, K., \& Hanawa, T. 1991, {\it
Astrophys. J.}, {\bf 367}, 213.

\bibitem[]{} Luminet, J.P. 1979, {\it Astr. Astrophys.}, 75, 228.

\bibitem[]{} Misner, C.W., Thorne, K.S., \& Wheeler, A.J. 1973, 
{\it Gravitation}, Freeman, San Fransisco.

\bibitem[]{} Mitsuda, K., Inoue, H., Koyama, K., Makishima, K.
et al. 1984, {\it Pub. Astr. Soc. Japan}, {\bf 36}, 741.

\bibitem[]{} Page, D.N., \& Thorne, K.S. 1974, {\it Astrophys. J.},
{\bf 191}, 499.

\bibitem[]{} Pandharipande, V.R. 1971, {\it Nucl. Phys.}, {\bf A178}, 
123.

\bibitem[]{} Popham, R., \& Sunyaev, R.A. 2000, {\tt astro-ph/0004017}

\bibitem[]{} Sahu, P.K., Basu, R., \& Datta, B. 1993, {\it Astrophys.
J.}, {\bf 416}, 267.

\bibitem[]{} Sunyaev, R. A. \& Shakura, N. I. 1986, {\it Sov. Ast. 
             Lett.}, {\bf 12}, 117

\bibitem[]{} Shimura, T., \& Takahara, F. 1995, {\it Astrophys. J.}, 
{\bf 445}, 780.

\bibitem[]{} Sun, W.-H., \& Malkan, M.A. 1989, {\it Astrophys. J.}, 
             {\bf 346}, 68

\bibitem[]{} Thampan, A.V., \& Datta, B. 1998, {\it Mon. Not. R. Astr. 
Soc.}, {\bf 297}, 570.

\bibitem[]{} Van der Klis, M. 2000, astro-ph/0001167.

\bibitem[]{} Wijnands, R., \& Van der Klis, M. 1998, {\it Nature},
{\bf 394}, 344.

\bibitem[]{} Walecka, J.D. 1974, {\it Ann. Phys.}, {\bf 83}, 491.

\end{thebibliography}
\end{document}